\documentclass[runningheads]{llncs}
\usepackage{graphicx}
\usepackage{footnote}

\usepackage[textsize=scriptsize,backgroundcolor=green]{todonotes} 

\begin{document}
\title{Devam vs. Tamam: 2018 Turkish Elections}
\author{Mucahid Kutlu\inst{1} \and
Kareem Darwish\inst{2} \and
Tamer Elsayed\inst{1}}
\authorrunning{Kutlu et al.}
\institute{Qatar University, Qatar \and
Qatar Computing Research Institute, HBKU, Qatar \\
\email{mucahidkutlu@qu.edu.qa}\\
\email{kdarwish@hbku.edu.qa} \\
\email{telsayed@qu.edu.qa}}
\maketitle              % typeset the header of the contribution
\begin{abstract}
On June 24, 2018, Turkey held a historical election, transforming its parliamentary system to a presidential one. 
One of the main questions for Turkish voters was whether to start this new political era with reelecting its long-time political leader Recep Tayyip Erdogan or not. In this paper, we analyzed 108M tweets posted in the two months leading to the election to understand the groups that supported or opposed Erdogan's reelection. We examined the most distinguishing hashtags and retweeted accounts for both groups. Our findings indicate strong polarization between both groups as they differ in terms of ideology, news sources they follow, and preferred TV entertainment.  

\end{abstract}

\section{Introduction}
On April 18, 2018, the Turkish president Recep Tayyip Erdogan announced early elections for the presidency and the parliament that would bring into force the constitutional changes that were approved by referendum on April 16, 2017. The constitutional changes would transform Turkey from a parliamentary system to a presidential system. With the office of the president enjoying significantly increased powers, these elections are highly consequential for Turkey. There were several presidential candidates representing the different Turkish political blocks such as conservatives, seculars, nationalists, and Kurds. Given the front runner status of the incumbent candidate Recep Tayyib Erdogan, we focus in this paper on those who support him and those who oppose him in favor of other candidates.

Ten days after the announcement of the elections, we started to crawl tweets by tracking keywords  relevant to the elections. As of June 23, 2018 (i.e., until the elections), we crawled 108M tweets. Using a semi-automatic labeling followed by a label propagation method, we labeled roughly 652.7K twitter users, of which 279.2K are pro-Erdogan and 373.5K are anti-Erdogan.  

Next, we analyzed the most distinguishing hashtags and retweeted accounts for each group. In our analysis, we found that anti- and pro-Erdogan groups follow different news media, have different demands from the governments and even watch different TV series, suggesting strong polarization that goes beyond political preferences. 

\section{Background}

Turkey has been enjoying democratic elections since 1950. While official results are announced by Supreme Election Council of Turkey\footnote{\url{http://www.ysk.gov.tr/}}, votes are counted by the members of political parties collectively, with each of them obtaining a copy of election results for each ballot box. Furthermore, elections and the counting process are also observed by international institutions. Therefore, the election process is usually considered secure and the results are accepted peacefully by all political parties. 

As a result of referendum on April 16, 2017, Turkey has made significant changes in its constitution, giving more power to the president. While the first election with the new constitution had been scheduled for 2019, the Turkish president Recep Tayyip Erdogan announced early elections to be held on June 24, 2018. In this election, Turkish citizens cast two different ballots, one for the presidency and the second for parliament.  Voter participation rate in both ballots was 86.24\%. % of voters for parliamentary and presidential elections are 86.24\% and  86.24\%, respectively.
 
Eight political parties participated in parliamentary elections, namely: 
\begin{enumerate}
\item The Justice and Development Party (AKP) which is Erdogan's party
\item The Nationalist Movement Party (MHP)
\item The Republican People's Party (CHP)
\item The People's Democratic Party (HDP)
\item The Iyi (good) Party (IYI)
\item The Saadet (felicity) Party (SP)
\item The Free Cause Party (Huda-Par)
\item The Motherland Party (VP)
\end{enumerate}

\begin{table}[!ht]
\begin{minipage}{\textwidth}
\label{tableparty}
  \begin{center}
  \begin{tabular}{  | l | p{6cm} | l | r |}
  \hline
  \textbf{Party} & \textbf{Description}  & \textbf{Alignment} &  \textbf{Election Results\footnote{\url{http://ysk.gov.tr/doc/dosyalar/docs/24Haziran2018/KesinSecimSonuclari/2018MV-96C.pdf}}} \\ \hline
  AKP & Erdogan's party & Public &  42.56\% \\ \hline
  MHP & Turkish nationalist party & Public &  11.1\% \\ \hline
  CHP & Secular party founded by Ataturk & Nation &  22.65\% \\ \hline
  IYI & Turkish nationalist party founded by mostly ex-members of MHP & Nation &  9.96\% \\ \hline
  SP & Islamist party & Nation &  1.34\% \\ \hline
  HDP & Secular Kurdish party that is mostly criticized for not distancing itself from the Kurdistan Workers Party (PKK), which is considered a terrorist group by Turkey, USA\footnote{\url{https://www.state.gov/j/ct/rls/other/des/123085.htm}} and many European countries. & None &  11.7\% \\ \hline
  Huda-Par & Kurdish Islamist party & None &  0.31\% \\ \hline
  VP & Left-wing nationalist party & None  &  0.23\% \\ \hline
   \end{tabular} 
  \end{center}
   \caption{\textmd{Political parties participated for parliamentary elections on June 24, 2018}. Abbreviations are used based on their official Turkish name.}
  \end{minipage}
 \end{table}

These parties offer a wide ideological spectrum for Turkish voters. For the first time in Turkish elections history, parties were also allowed to make alignments for parliamentary elections. AKP and MHP formed the ``Public's Alignment'', while CHP, IYI and SP formed the ``Nation's Alignment'', bringing parties with different ideological background together. 
\textbf{Table 1} % for some reason referencing is not working and it is writing Table 2, not 1
lists the political parties who participated in parliamentary elections and their election results.

In the presidential elections, there were 6 candidates, namely Recep Tayyip Erdogan (AKP), Muharrem Ince (CHP), Selahattin Demirtas (HDP), Meral Aksener (IYI), Temel Karamollaoglu (SP), and Dogu Perincek (VP). MHP and Huda-Par announced their support for Erdogan in the presidential election. \textbf{Table~\ref{table_presidential_elections}} lists the results of the presidential election.

\begin{table}[!ht]
\begin{minipage}{\textwidth}   
 \label{table_presidential_elections}
  \begin{center}
  \begin{tabular}{  | l | l | r |}
  \hline
  \textbf{Candidate Name} & \textbf{Party}   &  \textbf{Election Results\footnote{\url{http://ysk.gov.tr/doc/dosyalar/docs/24Haziran2018/KesinSecimSonuclari/2018CB-416D.pdf}}} \\ \hline
Recep Tayyip Erdogan   & AKP &  52.59\% \\ \hline
Muharrem Ince   & CHP &  30.64\% \\ \hline
Selahattin Demirtas   & HDP &  8.40\% \\ \hline
Meral Aksener   & IYI & 7.29\% \\ \hline
Temel Karamollaoglu   & SP &  0.89\%\\ \hline
Dogu Perincek   & VP & 0.20\%\\ \hline
\end{tabular}
  \end{center}
  \caption{\textmd{Presidential candidates and election results.}}
  \end{minipage}
 \end{table}

Other organizations with political influence include the Liberal Democratic Party, which is a liberal party, and the Hizmet Hareketi (service movement), which is a civil movement that the government refers to it as G{\"u}lenist Terror Organization (FET{\"O}) and accuses it of orchestrating the failed coup attempt in 2016. Also, Kemalist ideology, referring to principals of Mustafa Kemal Ataturk, is prevalent in Turkey, with adherents belonging to varying secular and nationalist parties.

\section{Data Collection and Labeling}
We collected tweets relating to Turkey and the elections starting on April 29, 2018 until June 23, 2018 -- the day before the election.  We tracked keywords related to the elections including political party names, candidate names, popular hashtags during this process (e.g., \#tamam and \#devam), famous political figures (e.g., Abdullah Gul, the former president of Turkey) and terms that may impact people's vote (e.g., economy, terrorism and others). We wrote keywords in Turkish with Turkish alphabet, which contains some additional letters that do not exist in the English alphabet (e.g., \c{c}, \u{g}, \"{u}).  Next, we  added versions of these keywords written strictly with English letters (e.g., ``Erdogan'' instead of ``Erdo\u{g}an'') allowing us to catch non-Turkish tweets about elections. 

\begin{figure*}
\begin{center}
\includegraphics[width=\linewidth]{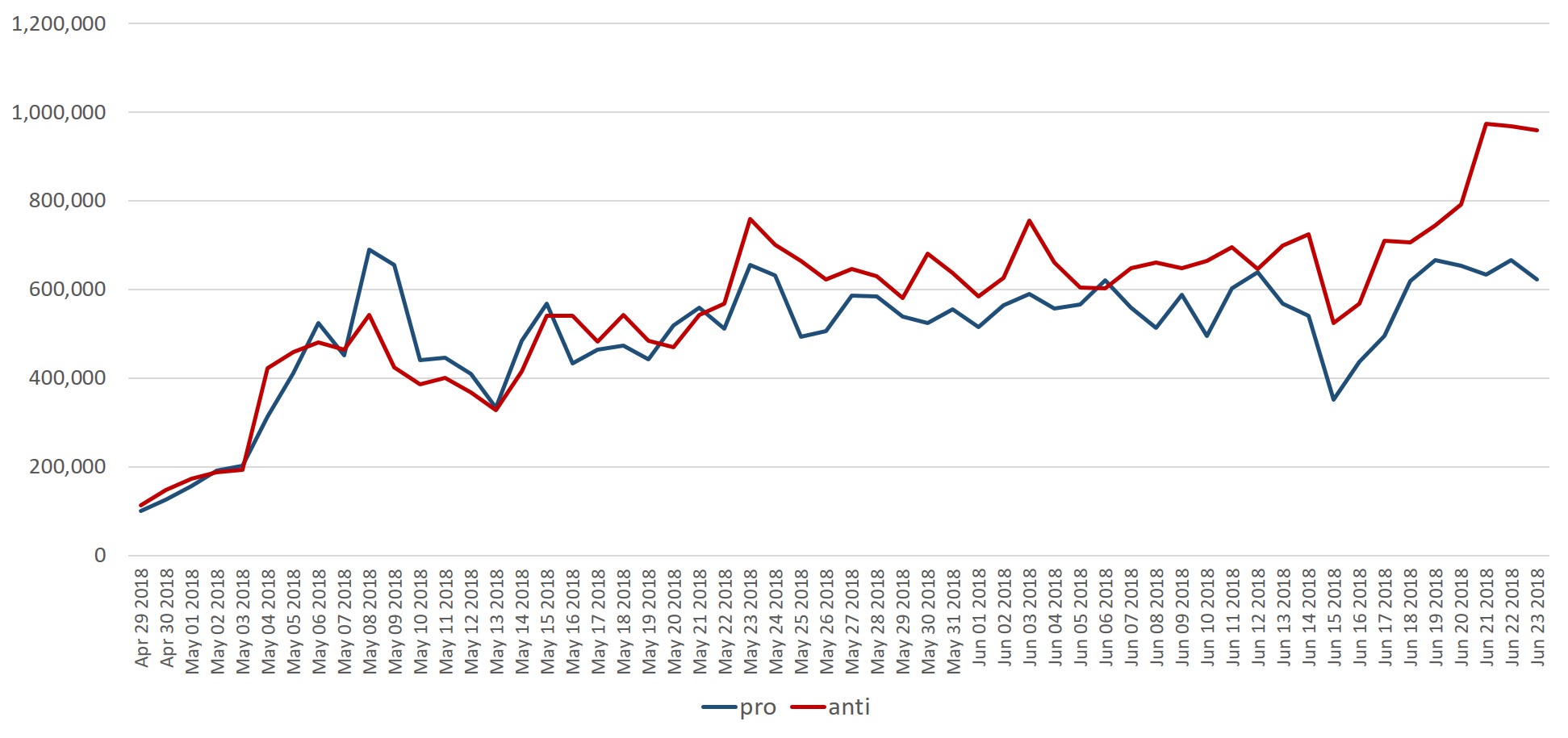}
\caption{Histogram of pro- and anti-Erdogan tweets}
\label{fig:histogram}
\end{center}
\end{figure*}

In the process we collected 108 million tweets.  Our first step was to label as many users by their stance as possible.  The labeling process was done in two steps, namely:
\begin{itemize}
\item \textbf{Manual labeling based on user names.} We assigned labels to users who explicitly specify their party affiliation in their Twitter name and screen name.  
%There are five major political parties in Turkey. The Justice and Development Party (AK Parti) and the Nationalist Movement Party (MHP) supported Erdogan in his reelection bid. The Republican People's Party (CHP), People's Democratic Party (HDP), and the Good Party (IYI) all fielded their own candidates. There are other smaller parties such as: Saadet party, which is an Islamist party that is generally anti-Erdogan; Huda-Par, which is a Kurdish Islamist party; and the Liberal Democratic Party, which is a liberal party. Another influential organizations is the Hizmet Hareketi (service movement), which is a Sufi movement and the government refers to it as G{\"u}lenist Terror Organisation (FET{\"O}) and accuses it of orchestrating the failed coup attempt in 2016.  
We made one simplifying assumption, namely that the supporter of a particular party would be supporting the candidate supported by their party. We extracted a list of users who use ``AKParti'' (official abbreviation of AKP), ``CHP'', ``HDP'', or ``IYI'' in their Twitter user or screen name.  We labeled the people who used ``AKParti'' as ``pro-Erdogan'', while the rest as ``anti-Erdogan''. Though ``MHP'' officially supported Erdogan in the election, we feared that the MHP supporters might not be universally supporting Erdogan. 

Further, the incumbent and front runner status of Erdogan caused his supporters to use the hashtag \#devam (meaning ```continue'') and his opponents regardless of their political affiliation to use the hashtag \#tamam (``enough''). Thus, we labeled users who had the hashtags \#devam or \#tamam in their profile description as supporting or opposing Erdogan.  Lastly, users who had the hashtag \#RTE (Recep Tayyip Erdogan) in their profile description were labeled as pro-Erdogan.  %In doing so, we were able to automatically tag 3,866 unique users.

While providing a political party name as a part of twitter user profile is a strong indication of supporting the respective party, we manually checked all extracted names to ensure the correctness of labels. For instance, we found that some users expressed that they are against a particular party in their user name instead of supporting it. Therefore, whenever we suspected that keywords we used for labeling were not indicative of their political view, we manually investigated the accounts and removed their labels if their political views were unclear. Table~\ref{table_aggregation_results} shows the number of our  manual labels.

\item \textbf{Label propagation.} %Using label propagation to label users based on their retweet behavior. 
Label propagation automatically labels users based on their retweet behavior \cite{magdy2016isisisnotislam}.  The intuition behind this method is that users that retweet the same tweets most likely share the same stances on the topics of the tweets. Given that many of the tweets in our collection were actually retweets or duplicates of other tweets, we labeled users who retweeted 10 or more tweets that were authored or retweeted by the pro- or anti- groups and no retweets from the other side as pro- or anti- respectively.  We iteratively performed such label propagation 11 times, which is when label propagation stopped labeling new accounts.  After the last iteration, we were able to label 652,729 users of which 279,181 were pro-Erdogan and authored 28,050,613 tweets and 373,548 were anti-Erdogan and authored 31,762,639 tweets. Figure~\ref{fig:histogram} shows the histogram of the pro- and anti-Erdogan tweets for the period of study.  As the figure shows, interest in the election has steadily increased on average as the election day approached.  There was a large drop in tweet activity around June 15, which corresponds to the end of Ramadan holiday (Ramazan Bayram{\i}) and is a public holiday.  Also, the anti-Erdogan group produced a greater number of tweets on most days, with the gap widening in the last few days of the campaign.
\end{itemize}

\begin{table}[!h]
 \label{table_aggregation_results}
  \begin{center}
  \begin{tabular}{ | l | c | }
  \hline
  \textbf{Supporters} & \textbf{No. of Labeled Users} \\ \hline
  pro-Erdogan  & 1,777\\ \hline
  anti-Erdogan w/out party affiliation & 2,134 \\  \hline
  pro-CHP & 833\\ \hline
  pro-IYI Party & 900 \\  \hline
  pro-HDP  & 357 \\ \hline
   \end{tabular}
  \end{center}
  \caption{\textmd{Number of Manually Labeled Users (3,866 Unique Users)}.}
 \end{table}

\section{Analysis}
Next we analyzed the data to ascertain the differences in interests and focus between both the pro- and anti-Erdogan groups as expressed by the hashtags that they use and the accounts that they retweet. Doing so can provide valuable insights into both groups \cite{darwish2017predicting,darwish2017trump}
% \item The common themes between both groups also as expressed using hashtags and retweeted accounts.
% \end{itemize}
To compute a relative importance score that accounts for the frequency of use by either group and the gap in use between both camps, we used the following formula:
\begin{equation}
score = tp * tpr * (1 - fpr)
\end{equation}

\noindent
where $tp$ and $fp$ are the number of times used by $1^{st}$ and $2^{nd}$ groups respectively, $tpr$ is the ratio within $1^{st}$ group $(tp/total_A)$, and $fpr$ is the ratio within $2^{nd}$ group $(fp/total_B)$.  Thus, the greater the usage within a group and the disparity between usages in both groups, the greater the score.  This formula is a variation on the Odds Ratio Numerator \cite{forman2003extensive}.

\textbf{Tables \ref{table:top50HashPro} and \ref{table:top50HashAnti}} list the top 50 hashtags and top 50 retweeted accounts for both groups organized by category.  \textbf{Table \ref{table:top50hashSummary}} summarizes both tables and provides percentages for each category.  As the Table \ref{table:top50hashSummary} shows, more than 51\% of hashtags used by the  pro-Erdogan group were in positive praise of Erdogan and AKP, Erdogan's  party, compared to 10.8\% of the hashtags dedicated to attacking other presidential candidates and their parties.  This ratio is roughly 5 to 1.  On the other side, roughly a third of the hashtags on the anti-Erdogan camp praised the other presidential candidates and their parties and roughly a fifth of the hashtags were attacking Erdogan and AKP. This ratio 1.7 to 1.  The large difference in ratio could be explained by the front runner status of Erdogan, where his proponents did not need to attack his opponents and instead needed to maintain a positive message. Though both groups discussed different social issues such as retirement, compulsory military service, and animal rights, the Palestinian issue featured prominently for the pro-Erdogan group. We proceeded to compare both groups based on the Palestinian issue by considering the union of the top 20 hashtags that mention Jerusalem (Turkish: Kud{\"u}s, English: Quds), Palestine (Turkish: Filistin, English: Palestine), or Nakba (meaning: catastrophe in Arabic) for both groups. As the comparison in \textbf{Table \ref{table:palestine}} shows, though both groups tweeted about the issue, the pro-Erdogan group were more than 6 times more likely to mention the Palestinian issue. Further, though some hashtags were common to both groups,  some were dominated by one group. For future work, we plan to see if this discrepancy between both groups are an artifact of the climate surrounding the election or the difference more systematic.

One of the things that caught our eye is the appearance of the names TV series in the top 50 hashtags.  Thus, we were curious if other TV series that users from both group refer to.  We obtained a list of 320 Turkish TV series that aired since 2014 to the present from Wikipedia\footnote{\url{https://en.wikipedia.org/wiki/List\_of\_Turkish\_television\_series}}.  \textbf{Table \ref{table:tvseries} }lists the series that were mentioned as hashtags with their full names at least 500 times and the percentage and count of mentions for both pro- and anti-Erdogan groups.  Two observations are quite apparent: 
\begin{itemize}
\item TV series preferences follow directly from political positions and show a very high level of polarization.  For example, the top 3 series in the table are mentioned more than 99\% of the times by anti-Erdogan users.
\item Anti-Erdogan users refer to TV series much more often than pro-Erdogan users suggesting that they are larger consumers of TV series.
\end{itemize}

We also examined the top retweeted accounts for anti- and pro-Erdogan accounts. We manually checked the accounts to identify their identity and political view. In order to identify political view of users, we checked their profile descriptions, most recent tweets and also profile pictures. For instance, users who use a picture of Ataturk for their profile or background image are assumed to be "Kemalist/Ataturkist". 

Concerning the top retweeted accounts (\textbf{Tables \ref{table:top50RTPro}, \ref{table:top50RTAnti}, and \ref{table:top5RTSummary}}), there are a few interesting patterns, namely:
\begin{itemize}
\item Official party accounts and ranking party member accounts accounted from the majority of retweets for the anti-Erdogan group (51.8\% -- 3.04 M tweets), while such accounts constituted roughly a third of retweeted accounts for the pro-Erdogan group (33.5\% -- 1.52 M tweets). % The number of retweets 3.04 M vs.\ 1.52M) 
\item The most retweeted journalist by pro-Erdogan accounts are either freelance journalists or work for online news portals. Only two journalist work at regular newspapers (Gunes and Posta). On the other hand, all top  retweeted journalists by anti-Erdogan accounts work (or have worked before) at popular TV channels or newspapers.  
\item Both groups retweet tweets from regular people who seems popular on social media.  However, we often observed strong ideological or political positions in these accounts.  For example, 4 (out of 9) of the top retweeted accounts by anti-Erdogan group use pictures of Ataturk in their profile suggesting that they are Kemalists.  Conversely, 2 (out of 6) of the top retweeted accounts by pro-Erdogan group use the picture of Erdogan in their profile. None of the regular accounts retweeted by pro-Erdogan group is identified as Kemalist.

\item Both groups retweeted accounts that don't have a clear owner such as \@RTECanli (Erdogan Live), which is dedicated to spreading positive news about Erdogan, and \@siyasifenomen (political phenomena), which is a dedicated anti-AKP account.  However, the percentage of such retweets for the pro-Erdogan group was 24\% compared to 10.7\% for anti-Erdogan group. % retweets tweets from users with nicknames more than anti-Erdogan group (839.8K vs.\ 628.8K).

\item Pro-Erdogan accounts were far more likely to retweet state media, while anti-Erdogan group rarely do so.  For example, \@TRTHaber was retweeted 70,711 and 1,289 times by the pro-Erdogan and anti-Erdogan groups respectively. Also, \@AnadoluAjansi was retweeted 61,623 and 1,232 times for both groups respectively. This is not surprising given that we observed the hashtag \#kapatgitsin (Turn off (state media)) being used by the anti-Erdogan group. On the other hand, the anti-Erdogan group was far more likely to retweet foreign news sources.  For example, sputnik\_TR was retweeted 64,767 and 4,907 times by the anti- and pro-Erdogan groups, respectively.  Similarly, the retweet counts for BBC Turkce were 47,314 and 2,376 for both groups, respectively. % retweets tweets from international media (i.e., BBC and Sputnik) while pro-Erdogan group does not. 

\item While Erdogan uses anti-FETO/PKK rhetoric in his political campaigns very frequently, only 2.7\% of retweets belong to accounts dedicated to the opposition of these groups.

% \item There is no account that have been retweeted by both groups.  
\end{itemize}

\section{Conclusion}
This paper presents an analysis of pro- and anti-Erdogan Twitter users using more than 108 million tweets related to the Turkish election that were collected from the period between April 29 and June 23, 2018.  We semi-automatically labeled a few thousands users based on their stance towards Erdogan and then we used label-propagation to tag more than 652 thousand users.  We examined the most distinguishing hashtags and retweeted accounts for both groups.  Our results show strong polarization that go beyond political preferences and extend to which channels people watch and which television series people follow.

\begin{table*}
\begin{center}
\begin{scriptsize}
\begin{tabular}{p{2cm}|l|r|p{7cm}}
Category & Hashtag & Freq. & Description \\ \hline
pro-Erdogan/  & devam & 286,463 &  "continue" \\
Establishment & VakitT{\"u}rkiyeVakti & 270,134 & AKP slogan ``It is Turkey time" \\
 & AKParti & 65,519 & AKP \\
 & T{\"u}rkiye{\c S}ahlan{\i}yor & 52,875 & "Turkey is developing” (AKP) \\
 & WeAreErdogan & 46,201 & ``We are Erdogan"  \\
 & Erdo{\u g}an & 45,625 & Erdogan \\
 & Erdo{\u g}anDemek & 39,573 & ``Erdogan means"   \\
 & MilletinAdam{\i}-MilletinAday{\i} & 27,812 & ``the man of the nation, the candidate of the nation"  \\
 & {\.I}ftiharVakti & 26,522 & ``time to be proud"  \\
 & VakitAnkaraVakti & 25,439 & ``It is Ankara's time" (AKP) \\
 & DurmakYokYolaDevam & 24,269 & ``no stopping"  \\
 & AKKad{\i}nlarSahada & 23,911 & ``women of AK Party are on the field", tweets reporting their activities for elections \\
 & VakitErzurumVakti & 22,959 & ``It is Erzurum's time" (AKP) \\
 & SendeDestekOl & 22,611 & ``you too support" pro-erdogan \\
 & Z{\"u}mr{\"u}d{\"u}Anka & 22,543 & ``emrald mythical bird" (Erdogan poem) \\
 & OyumAKPartiye & 22,117 & ``my vote to AKP" \\
 & Haz{\i}r{\i}zT{\"u}rkiye & 21,715 & ``Turkey, we are ready"  \\
 & B{\"u}y{\"u}kLider-{\.I}stanbulda & 21,656 & ``Great leader is in Istanbul"  \\
 & Vakit{\.I}stanbulVakti & 20,257 & ``It is Istanbul's time" (AKP) \\
 & VakitAdanaVakti & 20,245 & ``It is Adana's time" (AKP) \\
 \hline
TV shows & IVoteBtsbbmas & 225,714 & Music Awards \\ % (popular worldwide) \\
 & premiosmtvmiaw & 81,629 & Music Awards \\ %(popular worldwide) \\
 & MTVLakpopbts & 71,465 & Music Awards \\ %(popular worldwide) \\
 & Survivor2018 & 23,838 & TV show \\
 & MTVBrkpopbts & 21,003 & Music Awards \\ %(popular worldwide) \\
%  & Random &  &  \\
%  & canli & 67,223 & live (TV) \\
%  & sondak{\.I}ka & 44,413 & breaking news (TV) \\
 \hline
 Election related & 24Haziran & 93,854 & The election day \\
  \hline

Historical Events & 27May{\i}s	& 36,851 & "27 May", reminding the coup happened on 27 May, 1960. \\ \hline

Anti-Opposition  & KimlerKimlerleBeraber & 48,524 & ``who is with who" (anti CHP and HDP) \\
& ba{\c s}aramayacaks{\i}n{\i}z & 43,458 & ``you will not win" pro-erdogan slogan against anti-erdogan slogan ``we will win" \\
 & CHP & 36,691 & CHP \\
 & GelBakal{\i}mMuharrem & 24,457 & ``Come here Muharrem'' (anti-{\.I}nce) \\
 & hat{\i}rla & 22,437 & ``remember (how turkey has developed)'' \\ % for anti-Erdogan, ``remember the mistakes of Akparty" \\
 & Y{\i}k{\i}m{\.I}ttifak{\i} & 22,175 & ``destruction agreement'' (anti-opposition) \\
 \hline
 Issues & \textbf{\textit{Pro Palestine}} &  &  \\
 & WeStandForQuds & 38,203 & ``We stand with Jerusalem'' \\
 & Nakba70 & 31,340 & 70 years anniversary ``Nakba" (Palestinian catastrophe) \\
 & Kud{\"u}s & 26,483 & ``Jerusalem" \\
 & Kud{\"u}s{\.I}{\c c}inYeniKap{\i}day{\i}z & 23,339 & ``We are at Yenikapi for Jerusalem", calling for a rally in Yenikapi neighborhood of Istanbul \\
 & \textbf{Socio-economic} &  &  \\
 & Uzman{\c C}avu{\c s}larKimsesizMi & 33,530 & ``are sergeants alone" demanding better pay \\ % socio-economic status \\
 & Uzman{\c C}avu{\c s}larHakBekliyor & 21,626 & ``Sergeants are waiting for their rights" \\ % demanding better socio-economic status \\
 & 4clilerKa{\u g}{\i}t{\"U}zerinde4bli & 22,903 & Government officers are demanding better pay \\ % socio-economic status \\
 % & Agriculture engineers &  &  \\
 & gthb5bin & 25,745 & Demanding positions for agriculture engineers \\
 \hline
Positions & \textit{\textbf{Pro H{\"u}da-par}} &  &  \\
 & do{\u g}rusuH{\"u}dapar & 23,278 & ``H{\"u}da-par is the right one" \\
 & H{\"u}daparDiyorKi & 20,467 & ``H{\"u}da-par is saying that" \\
 & \textit{\textbf{Pro MHP}} &  &  \\
 &   MHP & 33,876 & MHP  \\ 
 & \textbf{\textit{Anti FET{\"O}}} &  &  \\
 & 26NisanFet{\"o}Kumpas{\i} & 20,607 & anti FETO \\
 \hline
\end{tabular}
\end{scriptsize}
\caption{Top 50 pro-Erdogan hashtags}
\label{table:top50HashPro}
\end{center}
\end{table*}

\begin{table*}
\begin{minipage}{\textwidth}   
\begin{center}
\begin{scriptsize}
\begin{tabular}{p{2cm}|l|r|p{7cm}}
Category & Hashtag & Freq. & Description \\ \hline
% Issues & Military service &  &  \\
Issues & BedelliAskerlik & 202,877 & military service -- pay not to go \\
% & Early retirement &  &  \\
 & EmeklilikteYa{\c s}aTak{\i}lanlar & 146,780 & demand for early retirement \\
 & EYT & 75,858 & demand for early retirement \\
% & Animal rights &  &  \\
 & Hayvana{\c S}iddetSu{\c c}tur & 15,652 & ``violence for animals is a crime'' pro animal rights \\
% & Currency devaluation &  &  \\
 & Dolar & 47,010 & currency devaluation \\
 & Euro & 24,978 & currency devaluation \\
\hline
Anti-Erdogan/  & Tamam & 232,829 & enough (anti-Erdogan) \\
Establishment & Erdo{\u g}an & 57,018 & Erdogan \\
 & KapatGitsin & 17,515 & ``Turn off'' (anti state media) \\
 & S{\i}k{\i}ld{\i}k & 17,146 & ``we are bored'' (tamam -- enough) \\
 & AKP & 27,287 & AKP \\
\hline
Pro-party  & \textbf{\textit{CHP}} &  &  \\
slogans/ & Muharrem{\.I}nce & 81,140 & CHP candidate \\
candidates  & Ba{\c s}araca\u{g}{\i}z & 77,139 & ``We will achieve(CHP) \\%& Muharrem{\.I}nce & 60,671 & CHP candidate \\
 & Hepimizin-Cumhurba{\c s}kan{\i} & 53,640 & ``President for All'' (CHP) \\
 & Muharrem{\.I}nce-HaberT{\"u}rkde & 36,009 & CHP candidate on Haber Turk \\
 & CHP & 29,060 & CHP \\
 & K{\i}l{\i}{\c c}daro\u{g}luNeS{\" o}yledi & 28,823 & ``what did Kilicdaroglu say'' Quotes from  the party leader \\
 & Muharrem{\.I}nce-CNNT{\"u}rkde & 26,485 & CHP candidate on CNN Turk \\
 & Muharrem{\.I}nce-21deNTVde & 23,473 & CHP candidate on NTV at 9:00 pm \\
 & BuSe{\c c}iminY{\i}ld{\i}z{\i}{\.I}nce & 21,565 & ``the star of this election is Ince'' CHP candidate \\
% & Muharrem{\.I}nce & 20,469 & CHP candidate \\ % merged with other muharrem ince. same meaning with different capital letters
 & Millet{\.I}{\c c}inGeliyoruz & 19,230 & ``Coming for the nation'' (CHP) \\
 & Muharrem{\.I}nce{\c C}{\"u}nk{\"u} & 15,663 & ``muharrem ince because'' CHP candidate \\
 & \textbf{\textit{HDP}} &  &  \\
 & demirta{\c s} & 71,986 & HDP candidate \\
 & hdp & 54,983 & HDP \\
 & SenleDe\u{g}i{\c s}ir & 35,878 & ``it changes with you'' (HDP) \\
 & \textbf{\textit{IYI}} &  &  \\
 & Y{\"u}z{\"u}n{\"u}G{\"u}ne{\c s}eD{\"o}nT{\"u}rkiye & 59,731 & ``Turn your face to the sun'' (IYI) \\
 & \textbf{\textit{Saadet Party}} &  &  \\
 & de\u{g}i{\c s}tir & 30,288 & ``change'' (Saadet) \\
\hline
TV shows & \textbf{\textit{Vatanim Sensin TV series}} &  &  \\
 & Vatan{\i}mSensin & 49,742 &  \\
 & hileon & 39,894 &  \\
 & Vatan{\i}mSensinVeda & 39,452 &  \\
 & MirayDaner & 15,690 &  \\
 & BoranKuzum & 15,314 &  \\
 & \textbf{\textit{Survivor TV show}} &  &  \\
 & survivor2018 & 29,514 &  \\
 & Hilmur & 20,561 &  \\
 & \textbf{\textit{Siyah \& Beyaz A{\c s}k}} &  &  \\
 & SiyahBeyazA{\c s}k & 28,623 & \\
 & Asfer & 18,451 &  \\
%  & Random &  &  \\
%  & canli & 16,421 & live (TV) \\
\hline
Election related & AdaylaraSoruyorum & 30,697 & ask the candidates (TV) \\
 & 24Haziran & 17,456 & election day \\
 & se{\c c}im2018 & 15,309 & 2018 elections \\
% &  izmir or {\.I}zmir & 52,877 & A city where CHP has the most  \\
% & {\.I}zmir & 24,202 &   \\
% &  izmir & 28,675 & A city where CHP has the most  \\
 \hline
Historical Events  & 19May{\i}s1919 & 26,205 & anniversary of Ataturk Youth and Sport Festival\footnote{the date assumed as Ataturk started the Independence War of Turkey} \\
  & Gezi5Ya{\c s}{\i}nda & 33,718 & 5 year anniversary of Gezi protests \\
\hline
News Media & GazeteS{\" o}zc{\" u} & 26,744 & Sozcu Newspaper, Secular Kemalist newspaper \\
\hline
\end{tabular}
\end{scriptsize}
\caption{Top 50 anti-Erdogan hashtags}
\label{table:top50HashAnti}
\end{center}
\end{minipage}
\end{table*}

\begin{table*}
\begin{center}
\begin{scriptsize}
\begin{tabular}{p{1.7cm}|p{2cm}|p{2cm}|r|p{6cm}}
Category & Account & Identity & Freq. & Description \\ \hline
AKP & RT\_Erdogan & Erdogan & 388,555 & Official account \\
 & Akparti & AKParti & 328,863 & Official account \\
 & 06melihgokcek & {\.I}brahim Melih G{\" o}k{\c c}ek & 165,588 &  ex-governor of Ankara \\
 & SavciSayan & Savc{\i} Sayan & 75,689 & Old CHP member  but currently an AKP member  \\
 & suleymansoylu & S{\" u}leyman Soylu & 70,429 &  minister of interior affairs \\
 & bayramsenocak & Bayram {\c s}enocak & 54,089 &  Head of Istanbul branch of AKP  \\
 & drbetulsayan & Dr. Bet{\"u}l Sayan Kaya & 44,006 & ex-minister \\
 & BurhanKuzu & Prof. Dr. Burhan Kuzu & 43,711 & co-founder of AKP  \\
 & mahirunal & Mahir {\"u}nal & 39,541 & Spokesman of  AKP  \\
\hline
Gov accounts & tcbestepe & Erdo{\u g}an & 222,061 & Official presidency  account\\
 & TC\_Basbakan & Binali Y{\i}ld{\i}r{\i}m & 101,918 & Prime minister \\
 & ikalin1 & {\.I}brahim Kal{\i}n & 45,013 & Erdogan's advisor and spokesman of Presidency \\ 
\hline
Journalists/ & fatihtezcan & Fatih Tezcan  & 162,739 &  Journalist, analyst \\
analysts  & omerturantv & {\" O}mer Turan & 88,083 & Journalist, analyst \\
 & Malikejder47 & Malik Ejder & 66,679 & Freelance journalist \\
 & turgayguler & Turgay G{\"u}ler & 64,025 & Journalist at Gunes Newspaper \\
 & HarunAlanoglu & Harun Alano{\u g}lu & 49,986 & Analyst \\
 & AsliAyDincer & Asl{\i} Ay Din{\c c}er & 47,638 & Journalist \\
 & abdullahciftcib & Abdullah {\c C}ift{\c c}i & 47,220 & Political strategist \\
 & Uzun-Abdurrahman & Abdurrahman Uzun & 41,568 & Journalist \\
 & nedimsener2010 & Nedim {\c s}ener & 40,734 & Journalist at Posta Newspaper \\
 & drlsmzwriter & Murat Sar{\i}ca & 36,683 & Journalist\\
\hline
Accounts & medyaadami & unknown & 246,125 & Freelance journalist \\
with & themarginale & unknown & 222,614 & Pro-Erdogan \\
nicknames & RTECanli & unknown & 82,839 & Pro-Erdogan \\
 & HurAktivist & unknown & 68,334 & Pro-Erdogan \\
 & Malazgirt\_Ruhu & unknown & 59,125 & Pro-Erdogan \\
 & maske3g & unknown & 51,779 & Pro-Erdogan \\
 & TheLaikYobaz & unknown & 45,739 & Pro-Erdogan \\
 & BestepeCB & unknown & 43,660 & Pro-Erdogan fan page \\
 & Enesicoo & Enes (no last name) & 37,815 &  Probably pro-Erdogan \\
 & UstAkilOyunlari & unknown &118,883 & Pro-Erdogan conspiracy theorist \\
 & siyasetcanli & unknown & 36,867 & Pro-Erdogan news media on twitter \\
 
 & enveryan & unknown  & 72,158 & Defines himself as Armenian (Turkish citizen) academic  though the identity is unclear\\
\hline
MHP & dbdevletbahceli & Devlet Bah{\c c}eli & 212,735 & MHP leader \\
 & MHP\_Bilgi & MHP & 94,213 & official  account \\
\hline
Regular Twitter users & GkhnKhrman & G{\"o}khan Kahraman  & 191,645 & popular pro-Erdogan account \\
 & slmhktn & Selami Haktan & 67,536 & popular pro-erdogan account \\
 & gelinNcikAKz & Cemile Ta{\c s}demir & 48,110 & Pro-erdogan \\
 & uguronal & U{\u g}ur {\"O}nal & 43,149 & Pro-erdogan  \\
 & AK\_suHandan & Handan Aksu & 38,239 & Pro-erdogan  \\
 & UmmetciSiyaset & Ayd{\i}n Binbo{\u g}a & 46,608 & Seems a pro-Palestine account \\
\hline

Anti-FETO/PKK  & GizliArsivTR & unknown & 84,413 & anti-FETO/PKK \\
accounts & FetoGercekleri & unknown & 37,261 & anti-FETO  \\ 
\hline
Media & trthaber & TRT & 70,711 & State media \\
 & yenisafak & Yeni {\c S}afak & 69,829 &  Yeni Safak news paper\\
 & stargazete & Star Gazete & 68,440 &  Star newspaper \\
 & anadoluajansi & Anadolu Ajans{\i} & 61,623 & State news agency \\
 & tvahaber & A Haber & 37,867 &  A Haber news channel \\
 & turkiyevakti & T{\" u}rkiye Vakti & 46,599 & AKP media \\
\hline
\end{tabular}
\end{scriptsize}
\caption{Top 50 retweeted accounts for pro-Erdogan group}
\label{table:top50RTPro}
\end{center}
\end{table*}

\begin{table*}
\begin{center}
\begin{scriptsize}
\begin{tabular}{p{2cm}|p{2.5cm}|l|r|p{7cm}}
Category & Account & Identity & Freq. & Description \\ \hline
CHP  & vekilince & Muharrem {\.I}nce & 947,866 & CHP candidate \\
 & erenerdemnet & Eren Erdem & 204,782 & ex-parliament member of CHP \\
 & tgmcelebi & Mehmet Ali {\c C}elebi & 156,808 & Leading CHP member \\
 & barisyarkadas & Bar{\i}{\c s} Yarkada{\c s} & 78,639 & Leading CHP member \\
 & ATuncayOzkan & Tuncay {\"O}zkan & 75,491 & CHP member \\
 & kilicdarogluk & Kemal K{\i}l{\i}{\c c}daro{\u g}lu & 72,027 & Party Leader \\
 & aykuterdogdu & Aykut Erdo{\u g}du & 66,769 & Leading CHP member \\
 & herkesicinCHP & CHP & 47,369 & official party account of CHP \\
 & senerabdullatif & Abd{\"u}llatif {\c S}ener & 44,567 & ex-AKP member but now in the parliament as a member of CHP \\
\hline
HDP  & hdpdemirtas & Selahattin Demirta{\c s} & 346,798 & HDP candidate \\
 & HDPgenelmerkezi & HDP & 117,731 & official account of HDP \\
 & sahmetsahmet & Ahmet {\c S}{\i}k & 76,526 & parliament member of HDP \\
 & barisatay & Bar{\i}{\c s} Atay & 47,871 & parliament member of HDP \\
 & ayhanbilgen & Ayhan Bilgen & 45,415 & Leading HDP member \\
\hline
IYI Party & meral\_aksener & Meral Ak{\c s}ener & 335,182 & IYI candidate \\
 & iyiparti & IYI & 72,829 & official party account \\
 & AytunCiray & Aytun {\c C}{\i}ray & 57,962 & Leading IYI member \\
\hline
Saadet Party & T\_Karamollaoglu & Temel Karamollao{\u g}lu & 116,953 & Saadet party  candidate\\
 & aliaktas7 & Ali Akta{\c s} & 47,648 & lawyer, supporter of Saadet party \\
\hline
Accounts with  & kacsaatoldunet & unknown & 382,711 & popular anti-Erdogan account \\
nicknames &  siyasifenomen & unknown & 138,302 & anti-Erdogan \\
& politikaloji & unknown & 54,339 & political view is unclear \\
& azyazarozyazarr & unknown  & 53,418 & Probably a kemalist account \\
\hline
Regular Twitter   & HasanGunaltay & Hasan G{\" u}naltay & 54,129 & Kemalist \\
accounts & esrapekdemir & Esra Pekdemir & 52,960 & Kemalist \\
 & F\_ISBASARAN1 & Feyzi {\.I}{\c s}ba{\c s}aran & 54,551 & former politician (AKP) but it seems he does not support AKP anymore \\
 & caapulcukiz & Simge Ekici & 49,901 & activist, a "GEZI" protestor \\
 & AtillaTasNet & Atilla Ta{\c s} & 72,623 & celebrity,  Anti-Erdogan  \\
 & M\_Selanik3 & Mustafa Selanik & 47,005 &  Anti-Erdogan \\
  & SunaVarol\_ & Suna Varol & 159,203 & Kemalist \\
 & errdemglr & Erdem & 90,345 & Kemalist \\
  & FidelOKAN & Fidel OKAN & 61,420 & lawyer/writer \\
\hline
Journalists & ismailsaymaz & {\.I}smail Saymaz & 113,835 & Journalist at Hurriyet newspaper \\
 & SedefKabas & Sedef Kaba{\c s} & 91,243 & Academic/journalist \\
 & fatihportakal & Fatih Portakal & 79,317 & Journalist at  Fox TV channel \\
 & ismaildukel & {\.I}smail D{\"u}kel & 75,873 & Journalist at  Halk TV channel \\
 & candundaradasi & Can D{\"u}ndar & 71,961 & journalist, ex-director of Cumhuriyet newspaper \\
 & ugurdundarsozcu & U{\u g}ur D{\"u}ndar & 60,593 & Journalist at Sozcu newspaper \\
 & mustafahos & Mustafa Ho{\c s} & 46,368 & Journalist, had worked at various news media \\
\hline
Liberal Democratic party & tokcem & Cem Toker & 84,051 & ex-leader of Liberal Democratic Party \\
\hline
Media & t24comtr & T24 & 78,632 & Online news media \\
 & halktvcomtr & Halk TV & 77,770 &  Halk TV channel \\
 & sputnik\_TR & Sputnik T{\"u}rkiye & 64,767 & Russian news media in Turkish \\
 & DikenComTr & Diken Gazete & 52,046 & Online news media \\
 & solhaberportali & Sol Haber & 48,796 & Online leftist news media \\
 & bbcturkce & BBC Turkey & 47,314 & BBC news media in Turkish \\
 & odatv & Oda TV & 46,440 & Online news media \\
 & gazetesozcu & S{\"o}zc{\"u} Gazetesi & 254,741 & Sozcu newspaper \\
 & cumhuriyetgzt & Cumhuriyet Gazete & 237,657 & Cumhuriyet newspaper \\
 & BirGun\_Gazetesi & BirG{\"u}n Gazetesi & 113,153 & Birgun  newspaper \\
\hline
\end{tabular}
\end{scriptsize}
\caption{Top 50 retweeted accounts for anti-Erdogan group}
\label{table:top50RTAnti}
\end{center}
\end{table*}

\begin{table*}
\begin{center}
\begin{small}
\begin{tabular}{l|r|r}
Category & Hashtag Count & Percentage \\
	& \multicolumn{2}{c}{Pro-Erdogan} \\ \hline
pro-Erdogan and Establishment & 1,108,446 & 50.8 \\
TV shows & 423,649 & 19.4 \\
Issues & 223,169 & 10.2 \\
Anti-Opposition & 197,742 & 9.1 \\
Positions towards others & 98,228 & 4.5 \\ 
Election related & 93,854 & 4.3 \\
Historical Events & 36,851 & 1.7 \\ \hline \hline
	& \multicolumn{2}{c}{Anti-Erdogan} \\ \hline
Pro-party slogans/mentions/candidate & 665,093 & 34.3 \\
Issues & 513,155 & 26.5 \\
Anti-Erdogan or Establishment & 351,795 & 18.2 \\
TV shows & 257,241 & 13.3 \\
Election related & 63,462 & 3.3 \\
Historical Events  &59,923 & 3.1  \\ 
News Media & 26,744 & 1.4 \\
%Ideological & 52,949 & 2.7 \\
%Adult content & 52,877 & 2.7 \\
\end{tabular}
\caption{Summary of 50 top pro- and anti- Erdogan hashtags}
\label{table:top50hashSummary}
\end{small}
\end{center}
\end{table*}

\begin{table*}
\begin{center}
\begin{small}
\begin{tabular}{l|r|r}
Category & RT Count & Percentage \\
	& \multicolumn{2}{c}{Pro-Erdogan} \\ \hline
% Gov \& Party official accounts & 1,041,397 & 23.0 \\
-- AKP & 1,210,471 & 26.7 \\
-- MHP & 306,948 & 6.8 \\
Both parties & 1,517,419 & 33.5 \\
Government Accounts & 368,992 & 8.1 \\ 
Journalists/analysts & 645,355 & 14.2 \\
Media & 355,069 & 7.8 \\
Regular Twitter accounts & 435,287 & 9.6 \\
Accounts with nicknames & 1,085,938 & 24 \\
% Party officials & 493,053 & 10.9 \\
Anti-FETO/PKK accounts & 121,674 & 2.7 \\\hline \hline
	& \multicolumn{2}{c}{Anti-Erdogan} \\ \hline
-- CHP party & 1,694,318 & 28.8 \\
-- HDP Party & 634,341 & 10.8 \\
-- IYI Party & 465,973 & 7.9 \\
-- Saadet Party & 164,601 & 2.8 \\
-- Liberal Democratic party & 84,051 & 1.4 \\
All parties & 3,043,284 & 51.8 \\
%Kemalists & 741,946 & 12.6 \\
Journalists & 539,190 & 9.2 \\
Media & 1,021,316 & 17.4 \\
Regular Twitter accounts & 642,137 & 10.9 \\
Accounts with nicknames & 628,770 & 10.7 \\
%Pro Erdogan account & 382,711 & 6.5 \\
%FETO account & 47,005 & 0.8 \\
\end{tabular}
\end{small}
\caption{Summary of 50 top pro- and anti- Erdogan retweeted accounts}
\label{table:top5RTSummary}
\end{center}
\end{table*}

\begin{table*}
\begin{center}
\begin{scriptsize}
\begin{tabular}{l|l|r|r|r}
hashtag & Translation & pro & anti & total \\ \hline
WeStandForQuds &  & 38,203 & 1,255 & 39,458 \\
Nakba70 &  & 31,340 & 6,127 & 37,467 \\
Kud{\"u}s & Jerusalem & 26,483 & 4,872 & 31,355 \\
Kud{\"u}s{\.I}{\c c}inYenikap{\i}day{\i}z & We are in Yenikapi (rally area) for Jerusalem & 23,339 & 1,670 & 25,009 \\
Filistin & Palestine & 14,778 & 3,721 & 18,499 \\
Kud{\"u}s{\.I}{\c c}inAyaktay{\i}z &  We stand for Jerusalem & 13,437 & 231 & 13,668 \\
KalbimizFilistinde & Our hearts are in Palestine & 7,551 & 3,313 & 10,864 \\
ZulmeLanetKud{\"u}seDestek & Cure the oppression, support Jerusalem   & 5,716 & 1,331 & 7,047 \\
Kud{\"u}s{\.I}{\c c}inSesVer & Raise your voice for Jerusalem  & 5,315 & 1,470 & 6,785 \\
Filistin{\.I}{\c c}inYenikap{\i}day{\i}z & We are in Yenikapi (rally area) for Palestine  & 5,568 & 1,184 & 6,752 \\
FilistinYaz4032yeG{\"o}nder & SMS 4032 to financially support Palestine  & 6,144 & 2 & 6,146 \\
FreePalestine &  & 1,475 & 778 & 2,253 \\
StandWithQuds &  & 1,989 & 77 & 2,066 \\
Palestine &  & 1,450 & 445 & 1,895 \\
Kud{\"u}s{\.I}slam{\i}nd{\i}r & Jerusalem is Islamic & 1,702 & 30 & 1,732 \\
Kud{\"u}s{\"u}nMi{\u g}feriRTE & RTE (Erdogan) is the helmet of Jerusalem   & 395 & 1,096 & 1,491 \\
BatmandaKud{\"u}sMitingi & Rally for Jerusalem in Batman (Turkish city)  & 1,170 & 61 & 1,231 \\
Filistin{\.I}{\c c}inBirlik &  Unity for  Palestine & 476 & 696 & 1,172 \\
Kud{\"u}seSelahaddinOl &  Be Salahaddin for Jerusalem & 874 & 3 & 877 \\
FilistineUmutOl &  Be the hope for Palestine & 743 & 72 & 815 \\
Kud{\"u}s{\.I}{\c c}inK{\i}yamVakti &  Time to uprise for Jerusalem & 803 & 1 & 804 \\
FreeQuds &  & 728 & 15 & 743 \\
Kud{\"u}seSahip{\c C}{\i}k & Support Jerusalem  & 520 & 121 & 641 \\
Nakba &  & 328 & 176 & 504 \\
QudsBelongsToIslam &  & 87 & 224 & 311 \\
SultanGaziKud{\"u}s{\.I}{\c c}inAyakta &  Sultan Gazi stands for Jerusalem & 38 & 177 & 215 \\
QudsDay4Return &  & 46 & 128 & 174 \\ \hline
\textbf{Sum} &  & 190,698 & 29,276 & 219,974 \\
\end{tabular}
\end{scriptsize}
\caption{The union of top 20 Palestine related hashtags for both pro- and anti- groups}
\label{table:palestine}
\end{center}
\end{table*}

\begin{table*}
\begin{center}
\begin{scriptsize}
\begin{tabular}{l| l|r|r|r|r|r}
Series & Type & pro \% & pro count & anti \% & anti count & sum \\ \hline
Vatan{\i}m Sensin &Drama, History, Romance & 0.7 & 345 & 99.3 & 50,629 & 50,974 \\
Siyah Beyaz A{\c s}k &  Action, Drama, Romance & 0.6 & 168 & 99.4 & 28,980 & 29,148 \\
Ad{\i}n{\i} Sen Koy & Drama, Romance & 0.2 & 32 & 99.8 & 13,093 & 13,125 \\
S{\"o}z &  Action, Drama, War & 3.8 & 276 & 96.3 & 7,084 & 7,360 \\
Fazilet Han{\i}m ve K{\i}zlar{\i} & Drama & 0.4 & 28 & 99.6 & 6,923 & 6,951 \\
Sen Anlat Karadeniz & Drama & 4.3 & 246 & 95.7 & 5,535 & 5,781 \\
{\c C}ukur &  Action, Crime, Thriller & 12.5 & 376 & 87.5 & 2,629 & 3,005 \\
Erkenci Ku{\c s} & Comedy, Drama & 0.5 & 13 & 99.5 & 2,495 & 2,508 \\
Dirili{\c s} Ertu\u{g}rul &  Action, Adventure, Drama, History & 79.7 & 1,695 & 20.3 & 432 & 2,127 \\
Ufak Tefek Cinayetler &  Action, Drama, Mystery & 34.2 & 638 & 65.8 & 1,225 & 1,863 \\
Payitaht Abd{\"u}lhamid &  Action, Drama, History & 93.3 & 1,218 & 6.7 & 88 & 1,306 \\
Cesur Y{\"u}rek & Action, Drama, Romance & 0.0 & 0 & 100.0 & 1,236 & 1,236 \\
Kad{\i}n &Drama & 39.0 & 445 & 61.0 & 695 & 1,140 \\
Kiral{\i}k A{\c s}k  & Comedy, Romance& 2.8 & 26 & 97.2 & 905 & 931 \\
Avlu &  Crime, Drama & 27.0 & 235 & 73.0 & 634 & 869 \\
Kara Para A{\c s}k &  Action, Crime, Drama & 0.9 & 8 & 99.1 & 840 & 848 \\
Yeter & Drama & 18.9 & 145 & 81.1 & 622 & 767 \\
Anne &  Drama & 84.7 & 508 & 15.3 & 92 & 600 \\
\end{tabular}
\caption{Most mentioned TV series in the tweets. Types of series are mostly retrieved from IMDB website.}
\label{table:tvseries}
\end{scriptsize}
\end{center}
\end{table*}

\bibliographystyle{splncs04}
\bibliography{ref.bib}

\begin{thebibliography}{1}
\providecommand{\url}[1]{\texttt{#1}}
\providecommand{\urlprefix}{URL }
\providecommand{\doi}[1]{https://doi.org/#1}

\bibitem{darwish2017predicting}
Darwish, K., Magdy, W., Rahimi, A., Baldwin, T., Abokhodair, N.: Predicting
  online islamophopic behavior after\# parisattacks. The Journal of Web Science
   \textbf{3}(1) (2017)

\bibitem{darwish2017trump}
Darwish, K., Magdy, W., Zanouda, T.: Trump vs. hillary: What went viral during
  the 2016 us presidential election. In: International Conference on Social
  Informatics. pp. 143--161. Springer (2017)

\bibitem{forman2003extensive}
Forman, G.: An extensive empirical study of feature selection metrics for text
  classification. Journal of machine learning research  \textbf{3}(Mar),
  1289--1305 (2003)

\bibitem{magdy2016isisisnotislam}
Magdy, W., Darwish, K., Abokhodair, N., Rahimi, A., Baldwin, T.: \#
  isisisnotislam or\# deportallmuslims?: Predicting unspoken views. In:
  Proceedings of the 8th ACM Conference on Web Science. pp. 95--106. ACM (2016)

\end{thebibliography}




% LNCS Homepage, \url{http://www.springer.com/lncs}. Last accessed 4
% \end{thebibliography}
\end{document}